\documentclass[sigconf]{acmart}
\usepackage{hyperref}
\usepackage{float}

\renewcommand\footnotetextcopyrightpermission[1]{} 
\begin{document}

\title{Fishy Cyber Attack Detection in Industrial Control Systems}
\subtitle{An approach based on sequence learning LSTM networks}

\author{Manikanta Reddy Dornala}{}
\affiliation{
	Indian Institute of Technology, Kanpur
}

\begin{abstract}
Cyber attacks have become serious threats to Industrial Control systems as well. It becomes important to develop a serious threat defense system against such vulnerabilities. For such process control systems, safety should also be assured apart from security. As unearthing vulnerabilities and patching them is not a feasible solution, these critical infrastructures need safeguards to prevent accidents,  both natural and artificial, that could potentially be hazardous. \\
Morita proposed an effective Zone division\cite{morita2013detection}, capable of evaluating remote and concealed attacks on the system, coupled with Principal Component Analysis. But the need to analyze the node that has been compromised and stopping any further damages, requires an automated technique.\\
Illustrating the basic ideas we'll simulate a simple Water plant. We propose a new automated approach based on Long Short Term Memory networks capable of detecting attacks and pin point the location of the breach.
\end{abstract}


\maketitle

\section{Introduction}
\label{sec:introduction}
In the age of Internet, cyber attacks have become a major threat. Until recently only private and information centered systems were breached. But now, cyber attacks are a threat to Industrial Systems as Well. Serious security vulnerabilities are patched in regular personal computers and commercial spaces, quite frequently but Industrial control systems are seldom fixed as these patches could lead to new conflicts in the system. 
\\
This opens up a wide space for attackers to sneak in. A prime example would be of Stuxnet, that sabotaged Iranian uranium enrichment facilities in 2010.

\subsection{The Stuxnet malware}
Stuxnet utilized existing vulnerabilities in the operating system, along with a good understanding of the PLCs to formulate a malware. The sole target of the malware was to manipulate the working of centrifuges that enriched Uranium. Such carefully crafted attacks cannot be prevented unless a perfect system is built. Instead, it is enough to detect such breaches and then assess the damage. Thus it is important to build a reliable security and safety mechanism to prevent against attacks like Stuxnet.

\subsection{Zone Based PCA}
Hashimoto proposed a method of securing the information system by dividing the network into "plural zones". By Zone Division \cite{hashimoto2013safety} the probability of detecting possible attacks and accidents is increased. 

Conjoined with PCA \cite{morita2013detection}, Zone Based PCA can analyze the relationship between the variables in \textit{plural zones} and detect any changes caused by potential concealed breaches and unintended accidents. There can many types of relationships between the variables that are analyzed differently by Zone-Based PCA. An on-board safety personnel is still required, who may notice the change in PCA variables and report an anomaly.

\subsection{Sequence Learning}
We propose a new approach based on sequence learning algorithms, to detect changes from regular working unlike relationships between variables. The neural network architecture, utilized is capable of learning how the normal functioning of the system looks like and detect if something has changed. 

In order to illustrate the working of the system, we simulate a simple water plant, that circulates hot water between two tanks.

\section{Simulated System}
\begin{figure}[H]
  \centering
  \includegraphics[width=\linewidth]{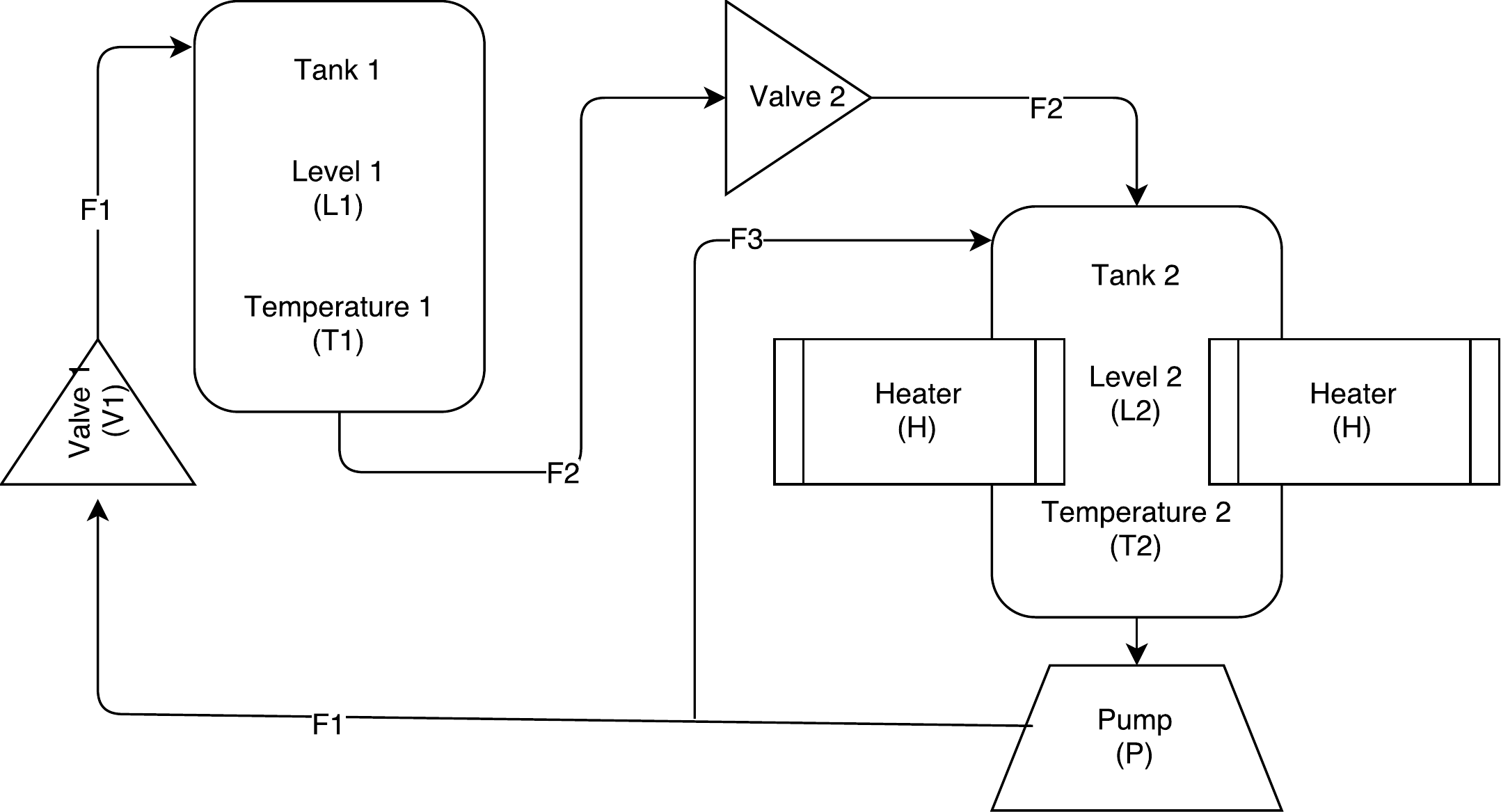}
  \caption{Simulated Water Plant}
  \label{fig:plant}
\end{figure}

\begin{figure}[H]
  \centering
   \includegraphics[width=0.8\linewidth]{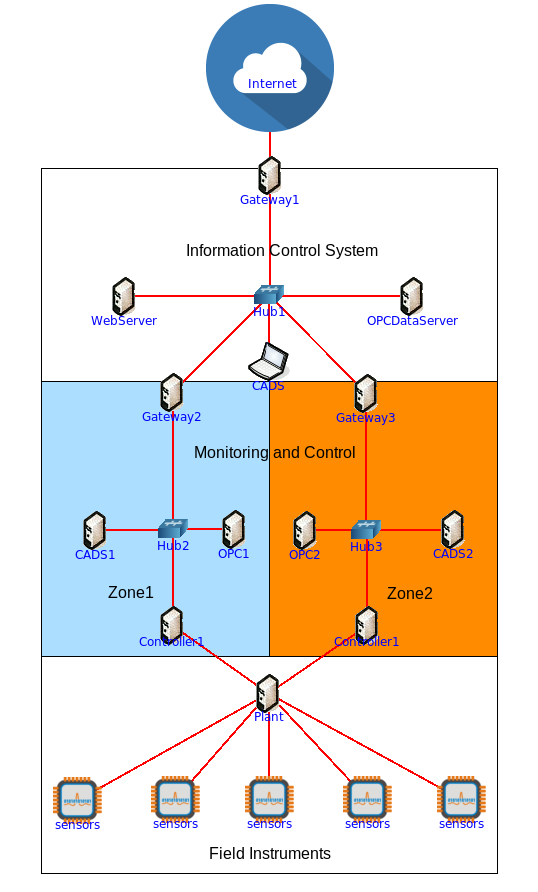}
  \caption{Information Flow Network of the Water Plant \newline OPC: OLE for Process Control,\newline CADS: Cyber Attack Detection System }
  \label{fig:cn}
\end{figure}

The simulation is a very basic version of the plant. In this system(Figure \ref{fig:plant}), water circulates between two tanks (Tank1 and Tank2) . The systems contain SCADA and other operators. The plant consists of many sensors and controls (Figure \ref{fig:cn}).

\subsection{Variables in the System}
L1 and L2 are levels of water in Tank1 and Tank2 respectively. Similarly, T1 and T2 measure the temperatures. The Heater (H) provides heat to increase the temperature of water in Tank2. The Pump (P) pumps the heated water into Tank1. Valves V1 and V2 are controlled to allow water to flow across them. Tank1 is assumed to radiate heat and cool down the water. 

We define two kinds of variables in our system. Process variables are the ones that are measured by sensors. Control Variables can be manipulated and change the state of the plant. \\

\textbf{Process Variables:} L1, L2, T1, T2, F1, F2, F3

\textbf{Control Variables:} V1, V2, H, P

\subsection{Network Configuration of the System}
The network is divided into two control zones,  such that the control of a control variable that directly affects a process variable is in a different zone. \\

\textbf{Zone 1:} L1, T1, V2, F2, H

\textbf{Zone 2:} L2, T2, V1, F1, F3, P
\\

A control variable in a zone cannot directly affect process variables in the same zone. In the current setup, L1 depends directly on the flow F1 controlled by V1. Therefore they are being separated. The rest of the division can be explained by similar logic fashion. 

H decides the amount of heat delivered into the system. There by it controls both T1 and T2, yet it is clubbed with T1 in Zone1 as there is more direct dependency between H and T2 as compared to H and T1. A formal method based on decidability matrices was presented by Hashimoto \cite{hashimoto2013safety}

\section{Zone Based PCA}
\begin{figure}[H]
  \centering
  \includegraphics[width=\linewidth]{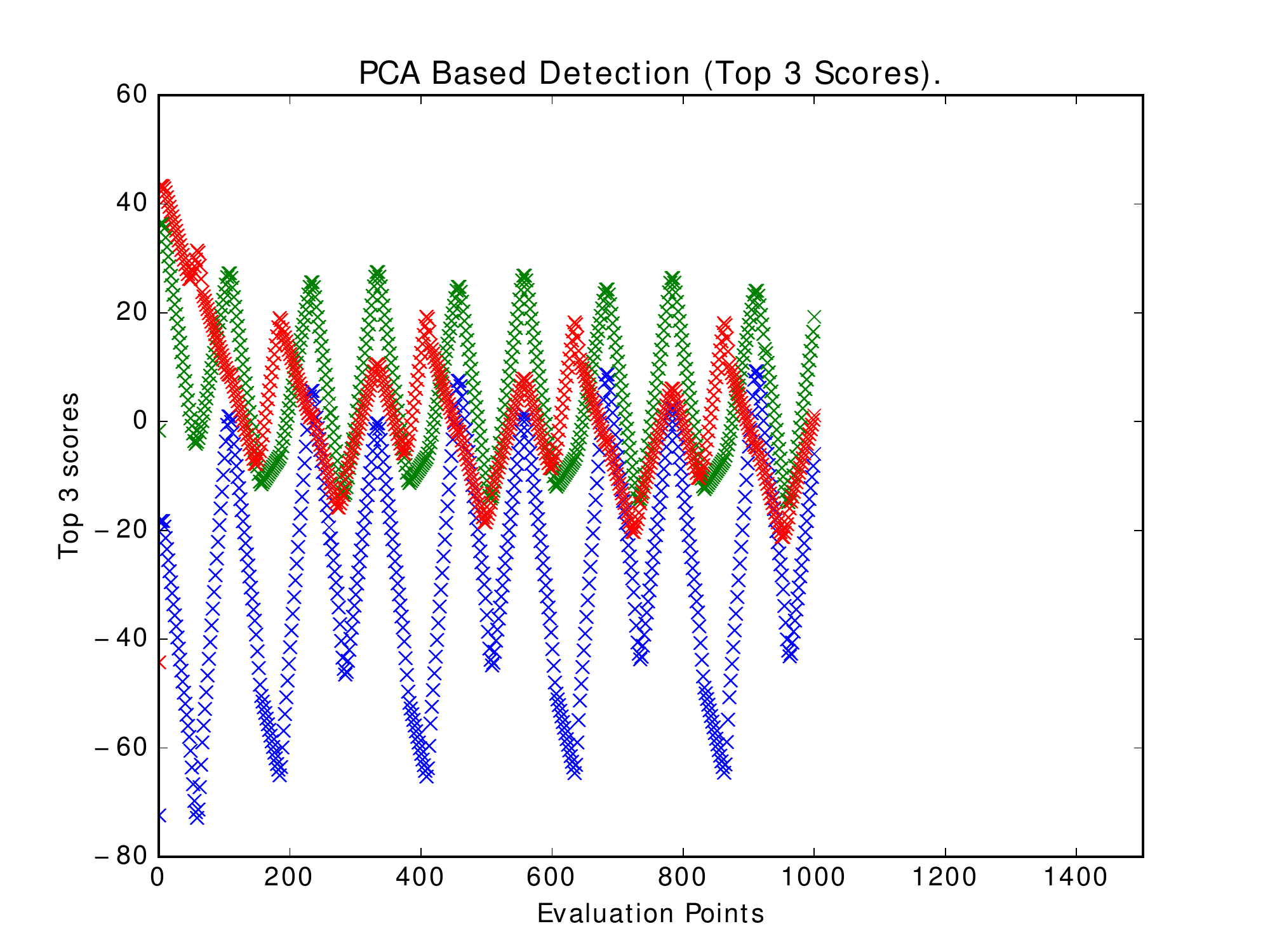}
  \caption{PCA projection of an un-compromised plant.}
  \label{fig:normal_pca}
\end{figure}

Principal Component analysis re-projects the data in the study into new space, with coordinates of high variance. Thus the variables with high variance can be maximally noted across them. PCA in a simple sense brings down the high dimensionality of the data to a smaller number. In our experiments, we've considered top 3 coordinates, in decreasing order of variance. Figure \ref{fig:normal_pca} shows this behavior of normal functioning.

The order of variance is as follows; blue, green, red in decreasing order.  The periodic nature of the plot is due to the way water is circulated. Now we'll simulate a data injection attack with concealment.

The attack manipulates Zone 2 and takes over the control of variables V1 and P. By setting them both to 0, the water in Tank1 doesn't decrease and the temperature of Tank1 increase to the extent where it depressurizes.

\begin{figure}[H]
  \centering
   \includegraphics[width=\linewidth]{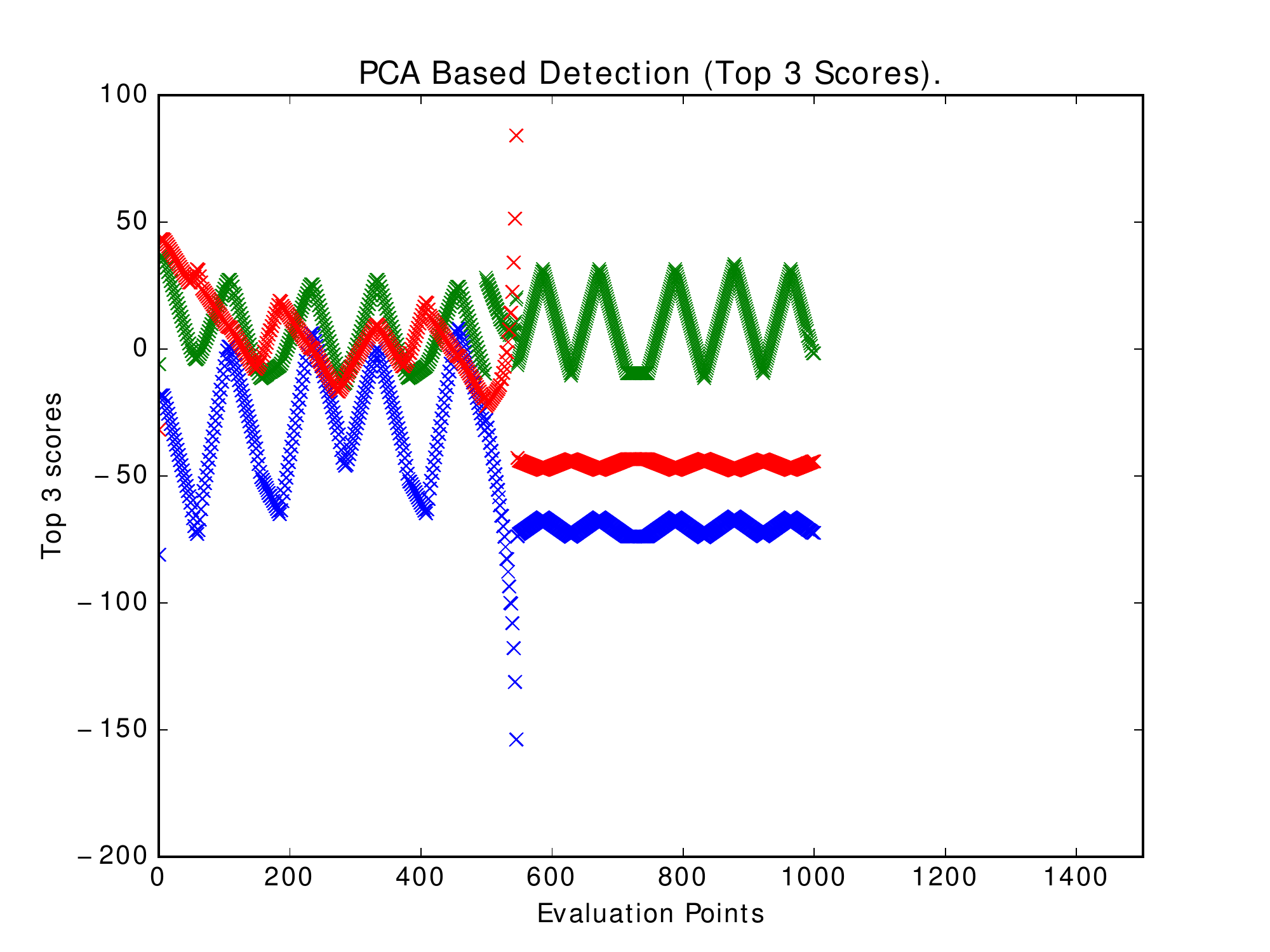}
  \caption{PCA projection of compromised plant.\newline \indent Zone 2 variables are remotely manipulated.}
  \label{fig:compromised_pca}
\end{figure}
Figure \ref{fig:compromised_pca} shows the dynamics of this process. The attack begins at simulation point 500.

There is drastic change in the behavior of the projection in blue. Any on-board personnel can realize such a change of high magnitude and shut down the systems if necessary.

It becomes difficult with increasing types of manipulation to identify what kind of change in the systems produces a particular kind of behavior in the PCA projection. 

\section{Zone based LSTM network}
The PCA method evaluates relationships between variables and raises alarm when the relationships or the dependencies between the variables changes. The relationships can change in many different ways when the number of variables involved is very large.

Instead of understanding the dependency between them, we focus on differentiating the modes of running of the plant as a whole. A normal working state is very different from a compromised state. As the systems tend to repetitive work, it is not hard to notice that there is a pattern to the way an un-compromised plant produces data. 

In order to achieve this, we use sequence learning LSTM networks in our method.

\subsection{Long Short Term Memory}

\begin{figure}[H]

  \centering
  \includegraphics[width=\linewidth]{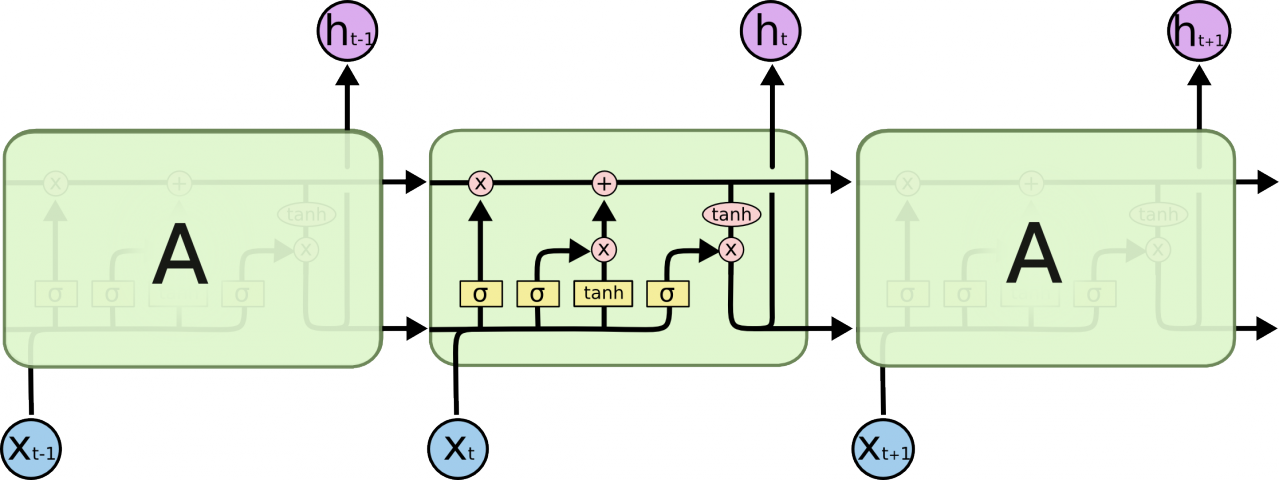}
  \caption{%
  Long Short Term Memory, LSTM, Units in action.
  Img ref: \href{http://colah.github.io/posts/2015-08-Understanding-LSTMs/}{colah}
}
  \label{fig:lstm}
\end{figure}
LSTM\cite{hochreiter1997long} is a form of Recurrent Neural Network, it remembers what's important and forgets the trivial things. It is capable of learning patterns in data and identify those that don't match. 

LSTM achieves this by what are termed as memory control gates in the unit. There are different gates to control and filter data. The gates decide what kind of variables at a simulation point are important. By doing so repeatedly, it learns the values of the variables to look for to understand the sequence pattern. It then assigns a score , that signifies how strongly it believes in the pattern.

We train an LSTM network with a logged data of the normal functioning of plant, and use it to give a confidence score to the previous 50 points at every evaluation point. This enables us to know if the current sequence of data generated is in high correlation with data generated during normal functioning.

\begin{figure}[H]

  \centering
  \includegraphics[width=\linewidth]{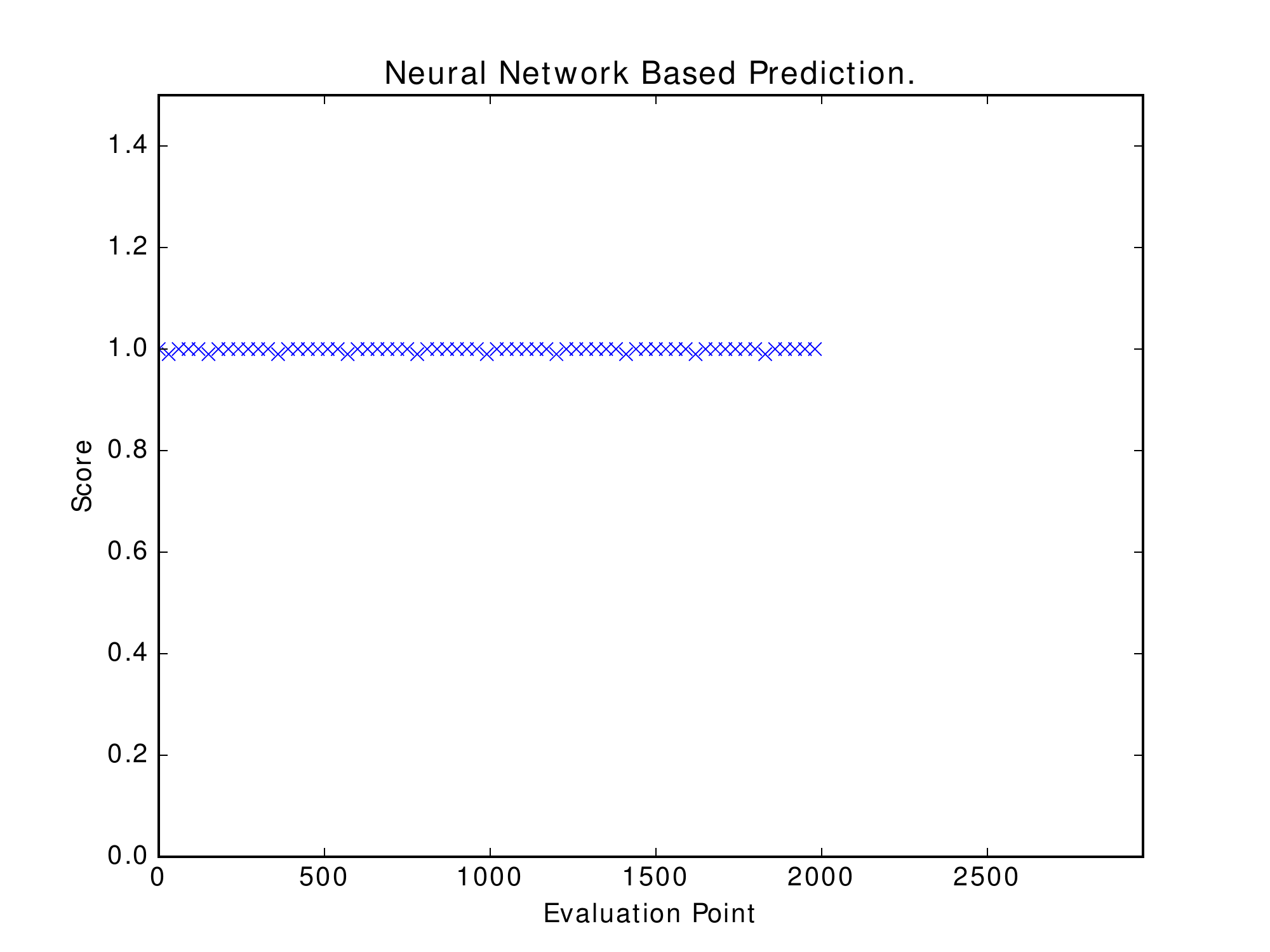}
  \caption{LSTM score of an un-compromised plant.}
  \label{fig:normal_nn}
\end{figure}

Normal functioning ensures a score match near close to 1, which is expected.

\begin{figure}[H]
  \centering
  \includegraphics[width=\linewidth]{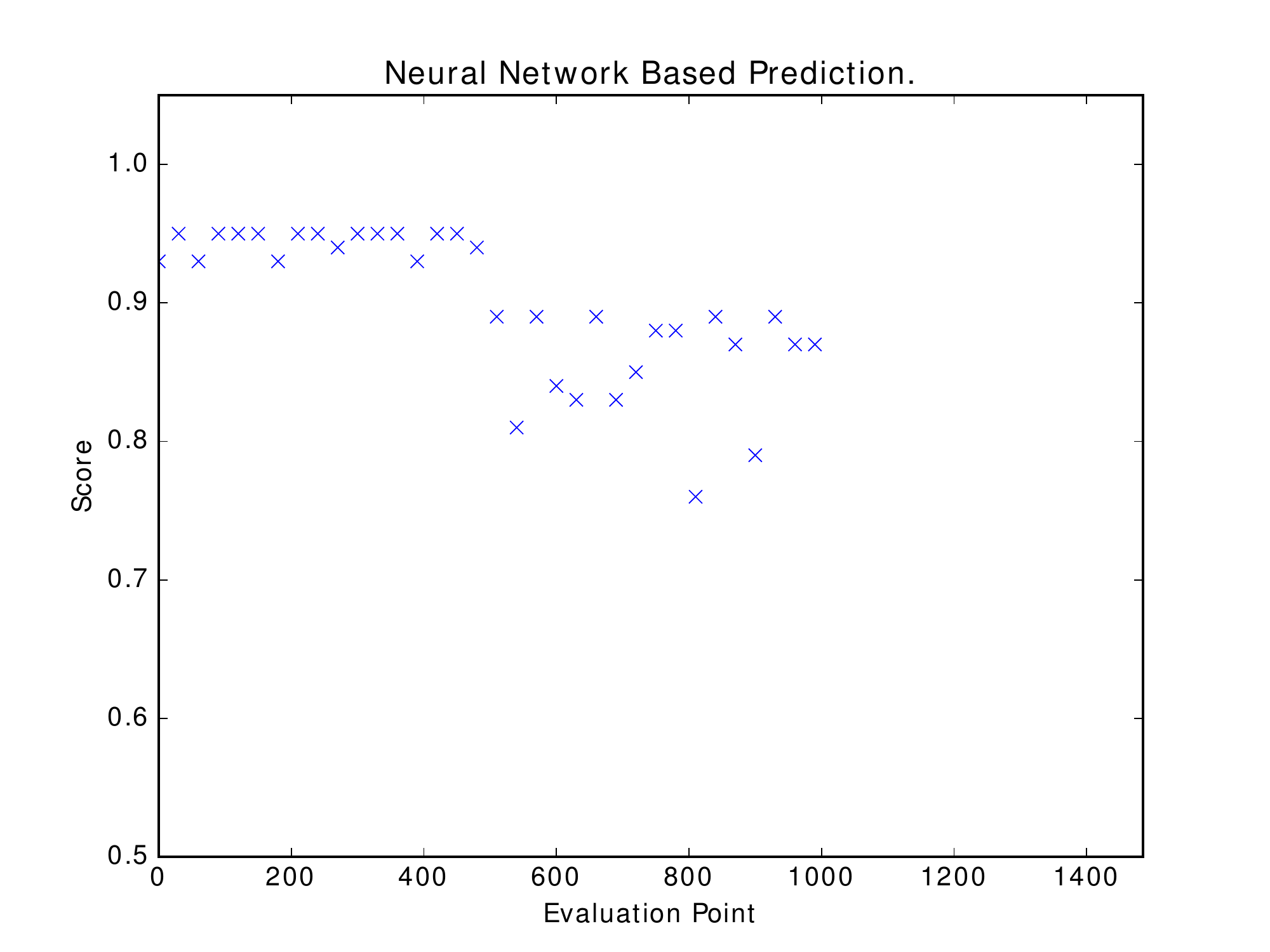}
  \caption{LSTM score of compromised plant.\newline \indent Zone 2 variables are remotely manipulated and P is set to 0.}
  \label{fig:compromised_nn00}
\end{figure}

In Figure \ref{fig:compromised_nn00} we compromise zone2 in a similar fashion. Both P and V1 are set to 0 here.

It is to be noted that when the zone is compromised, the running of the plant produces a different pattern of data. Then the LSTM network gives us a correlation score with the actual pattern of functioning, which in this case, when P is set to 0 averages around a value of 0.85.

\begin{figure}[H]
  \centering
  \includegraphics[width=\linewidth]{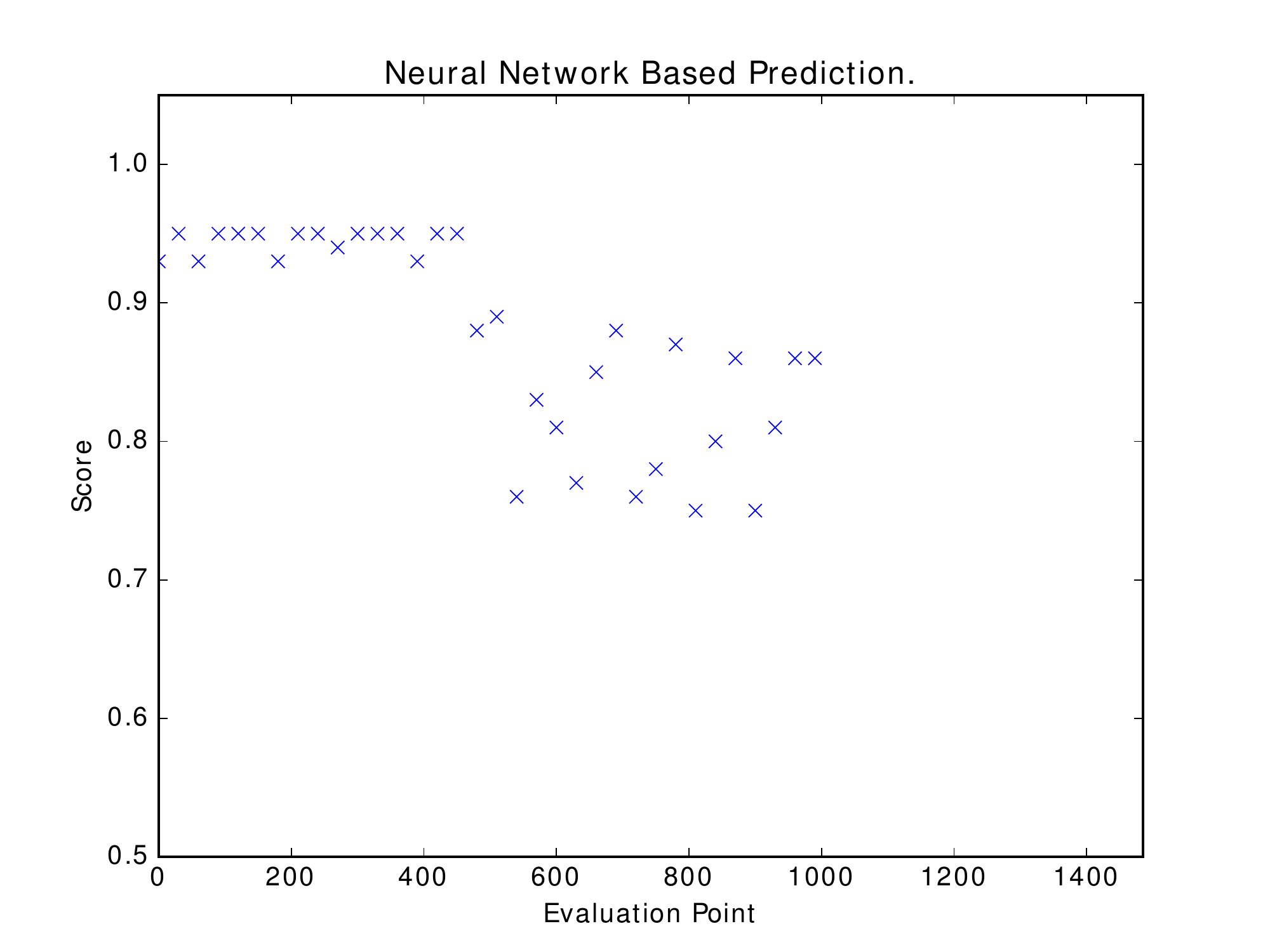}
  \caption{LSTM score of compromised plant.\newline \indent Zone 2 variables are remotely manipulated and P is set to 1.}
  \label{fig:compromised_nn11}
\end{figure}

When the variable P is set to 1 (Figure \ref{fig:compromised_nn11}), the pattern of execution produces scores averaging around 0.8.

By thresholding at different values we can easily identify which variable the attacker has modified. We can also detect the zone which has been compromised.

\noindent A plausible set of thresholds to detect the variable could be. 
\\
\noindent\textbf{P = 0:} score $\in$ (0.8, 0.9)

\noindent\textbf{P = 1:} score $\in$ (0.75, 0.85)
\section{Conclusion}
Attack Detection is of paramount importance as opposed to unearthing vulnerabilities and fixing them, in Industrial Systems due to high latency in patching their systems. The proposed method for intrusion detection, based on LSTM is also capable of diagnosing the attack for points of failure. Sequence-based learning and anomaly detection have an advantage over PCA-based methods in this regard. This approach shows an example of interdisciplinary work on implementing machine learning technologies for tackling the problems of Industrial Systems and Networks.

\end{document}